\documentclass[twocolumn]{aastex631}
\usepackage{graphicx} 
\usepackage{amsmath}

\begin{document}

\title{Gravitational-Wave Parameter Estimation in non-Gaussian noise\\ using Score-Based Likelihood Characterization}

\author{Ronan Legin}
\affiliation{Ciela - Montreal Institute for Astrophysical Data Analysis and Machine Learning, Montréal, QC, Canada}
\affiliation{Department of Physics, Université de Montréal, Montréal, QC, Canada}
\affiliation{Mila - Quebec Artificial Intelligence Institute, Montréal, QC, Canada}

\author{Maximiliano Isi}
\affiliation{Center for Computational Astrophysics, Flatiron Institute, New York, NY, USA}

\author{Kaze W. K. Wong}
\affiliation{Center for Computational Astrophysics, Flatiron Institute, New York, NY, USA}
\affiliation{Department of Applied Mathematics and Statistics, Johns Hopkins University, Baltimore, MD, USA}

\author{Yashar Hezaveh}
\affiliation{Ciela - Montreal Institute for Astrophysical Data Analysis and Machine Learning, Montréal, QC, Canada}
\affiliation{Department of Physics, Université de Montréal, Montréal, QC, Canada}
\affiliation{Mila - Quebec Artificial Intelligence Institute, Montréal, QC, Canada}
\affiliation{Center for Computational Astrophysics, Flatiron Institute, New York, NY, USA}
\affiliation{Trottier Space Institute, Montréal, QC, Canada}
\affiliation{Perimeter Institute for Theoretical Physics, Waterloo, ON, Canada}

\author{Laurence Perreault-Levasseur}
\affiliation{Ciela - Montreal Institute for Astrophysical Data Analysis and Machine Learning, Montréal, QC, Canada}
\affiliation{Department of Physics, Université de Montréal, Montréal, QC, Canada}
\affiliation{Mila - Quebec Artificial Intelligence Institute, Montréal, QC, Canada}
\affiliation{Center for Computational Astrophysics, Flatiron Institute, New York, NY, USA}
\affiliation{Trottier Space Institute, Montréal, QC, Canada}
\affiliation{Perimeter Institute for Theoretical Physics, Waterloo, ON, Canada}



\begin{abstract}
Gravitational-wave (GW) parameter estimation typically assumes that instrumental noise is Gaussian and stationary. Obvious departures from this idealization are typically handled on a case-by-case basis, e.g., through bespoke procedures to ``clean'' non-Gaussian noise transients (glitches), as was famously the case for the GW170817 neutron-star binary. Although effective, manipulating the data in this way can introduce biases in the inference of key astrophysical properties, like binary precession, and compound in unpredictable ways when combining multiple observations; alternative procedures free of the same biases, like joint inference of noise and signal properties, have so far proved too computationally expensive to execute at scale. Here we take a different approach: rather than explicitly modeling individual non-Gaussianities to then apply the traditional GW likelihood, we seek to learn the true distribution of instrumental noise without presuming Gaussianity and stationarity in the first place. Assuming only noise additivity, we employ score-based diffusion models to learn an empirical noise distribution directly from detector data and then combine it with a deterministic waveform model to provide an unbiased estimate of the likelihood function. We validate the method by performing inference on a subset of GW parameters from 400 mock observations, containing real LIGO noise from either the Livingston or Hanford detectors. We show that the proposed method can recover the true parameters even in the presence of loud glitches, and that the inference is unbiased over a population of signals without applying any cleaning to the data. This work provides a promising avenue for extracting unbiased source properties in future GW observations over the coming decade.
\end{abstract}



\section{Introduction} \label{sec:intro}

Gravitational-wave (GW) detectors, like LIGO \citep{LIGOScientific:2014pky}, Virgo \citep{VIRGO:2014yos} and KAGRA \citep{Aso:2013eba}, record noisy time series encoding astrophysically-valuable signals from black hole and neutron star collisions \citep{LIGOScientific:2016aoc,LIGOScientific:2017vwq,LIGOScientific:2021djp}. Extracting science from these data requires using Bayesian inference to estimate the source parameters,  \citep[e.g., black hole masses and spins, ][]{Veitch:2014wba,Ashton:2018jfp,Romero-Shaw:2020owr,Biwer:2018osg,Wong:2023lgb}. As in many other areas of astronomy, GW parameter estimation has traditionally assumed that instrumental noise is Gaussian and stationary \citep{LIGOScientific:2019hgc}. This idealization is approximately justified for short segments of data by the central limit theorem \citep{Berry:2014jja}, and has the additional advantage of having a tractable and relatively inexpensive likelihood.

In reality, however, the statistics of noise often deviate significantly from a stationary Gaussian process: the instruments evolve over time, and the data are contaminated by both transient non-Gaussian excursions (``glitches'') and the nonlinear evolution of narrow spectral features (``lines'') \citep{LIGOScientific:2016gtq,Abbott:2016xvh,aLIGO:2020wna}; in principle, the data are also contaminated by subthreshold astrophysical signals \citep{LIGOScientific:2021usb}.
Since no generative models exist for most of these contaminants, they are usually handled by bespoke treatment of the affected data segments \citep{Cornish:2014kda,Cornish:2020dwh,Davis:2018yrz,Chatziioannou:2021ezd,Davis:2022ird,Hourihane:2022doe,Udall:2022vkv,Ashton:2022ztk}, which can be computationally expensive and/or result in biases.
The most prominent example of this took place for the first binary neutron star merger \citep{LIGOScientific:2017vwq}, which was famously contaminated by a loud glitch in one of the LIGO detectors \citep{Pankow:2018qpo}. Non-Gaussian contamination has also interfered with other key scientific targets, such as identifying precession or anti-aligned spins in a black-hole binary \citep{Hannam:2021pit,Payne:2022spz,Macas:2023wiw,Udall:2024ovp}. In recent observing runs, approximately 20\% of detections have required mitigation of non-Gaussianities, highlighting the need for statistically robust, reproducible, and scalable procedures for addressing departures from the stationary-Gaussian likelihood \citep{Davis:2022ird}.

In this work, we showcase a new framework for GW parameter estimation without assuming stationary Gaussian noise, and without sacrificing the advantages of deterministic signal models.
This makes it possible to analyze signals contaminated by noise artifacts without the need for special treatment, and reduces biases in the analysis of large collections of signals which would otherwise be sensitive even to small departures from the Gaussian idealization \citep{Heinzel:2023vkq}. Our method differs from existing approaches, which attempt to either (1) directly model the non-Gaussianities as mentioned above, or (2) generalize the likelihood through simulation-based inference by giving up the known generative signal model \citep{2021PhRvL.127x1103D, 2024PhRvD.109f4056L,Dax:2024mcn,Xiong:2024gpx,Raymond:2024xzj}.

We adapt the method in \citet{legin}, score-based likelihood characterization (SLIC), which  was originally developed to model noise in telescope image data:
 we train a score-based diffusion model on samples of LIGO noise to learn the gradient of the log probability density of the noise distribution \citep{Legin2023icml}. Combined with the Jacobian of the waveform model, we can construct the likelihood function and use gradient-based samplers to draw from the posterior distribution of GW parameters. In this way, we can directly estimate the true likelihood of detector noise, simultaneously addressing issues of non-Gaussianity and non-stationarity while avoiding the need to subtract best-fit estimates of glitches in the data (which induces biases) or to co-model the glitch with the signal (which is prohibitively expensive when analyzing entire catalogs) and, crucially, without relinquishing deterministic signal templates.

We evaluate the method by inferring the posterior distribution from 400 mock observations of simulated gravitational wave signals containing real noise from the LIGO Livingston and Hanford detectors. Coverage probability tests show no evidence of bias in our posteriors.

The paper is structured as follows: In Section \ref{sec:methods}, we describe the methodology for learning the noise distribution and performing inference. Section \ref{sec:data} outlines the data used to model the noise distribution. Sections \ref{sec:neural_network}--\ref{sec:inference} cover the neural network architecture and training, the waveform (forward) model and the inference setup. In Section \ref{sec:results} we present our results, and in Section \ref{sec:discussion} we conclude.

\section{Methods}
\label{sec:methods}

\subsection{Problem Setup}
\label{sec:setup}

The goal of parameter estimation (PE) is to obtain samples from the posterior distribution $p(\boldsymbol{\theta}|\boldsymbol{d})$, where $\boldsymbol{\theta}$ represents the parameters of interest, such as the masses and spins of the GW binary, and $\boldsymbol{d}$ is the time series data recorded by one or more instruments.

Using Bayes's theorem, the posterior can be rewritten as $p(\boldsymbol{\theta}|\boldsymbol{d}) \propto p(\boldsymbol{d}|\boldsymbol{\theta})\, p(\boldsymbol{\theta})$, where $p(\boldsymbol{d}|\boldsymbol{\theta})$ is the likelihood and $p(\boldsymbol{\theta})$ is some prior distribution. The likelihood function contains all information about the data generation process; for GWs, this process can be expressed as
\begin{equation}
\label{eq:data_gen}
    \boldsymbol{d} = \boldsymbol{h}(\boldsymbol{\theta}) + \boldsymbol{n}\, ,
\end{equation}
where $\boldsymbol{n}$ represents the instrumental noise and $\boldsymbol{h}$ is the waveform (forward) model. If $ \boldsymbol{h}(\boldsymbol{\theta})$ is deterministic, then the probability of observing $\boldsymbol{d}$ given $ \boldsymbol{\theta} $ is equivalent to the probability of observing a given residual, $ \boldsymbol{d} - \boldsymbol{h}(\boldsymbol{\theta})$,
\begin{equation} \label{eq:likelihood}
p(\boldsymbol{d} \mid \boldsymbol{\theta}) = p(\boldsymbol{d} - \boldsymbol{h}(\boldsymbol{\theta}))\, .
\end{equation}
From Eq.~\eqref{eq:data_gen}, it follows that $p(\boldsymbol{n}) = p(\boldsymbol{d} - \boldsymbol{h}(\boldsymbol{\theta}))$, i.e., the likelihood function $p(\boldsymbol{d} \mid \boldsymbol{\theta})$ is equivalent to the probability distribution of the noise, $p(\boldsymbol{n})$.

Traditional PE assumes instrumental noise is Gaussian and stationary, with a known power-spectral-density (PSD), so that $p(\mathbf{n})$ has a simple closed form depending only on the second moment of the residual \citep{LIGOScientific:2019hgc}.
This introduces biases in the inference if the noise distribution is not strictly Gaussian and stationary with the assumed PSD. Therefore, we instead propose to machine-learn the noise distribution $p(\boldsymbol{n})$ directly from real data. We can then compute the likelihood by evaluating $p(\boldsymbol{n})$ at the residuals $\boldsymbol{d} - \boldsymbol{h}(\boldsymbol{\theta})$ for any given $\boldsymbol{\theta}$, using any suitable waveform model.

\subsection{SLIC Framework}
\label{sec:slic}

As a distribution, the likelihood is defined over the space of possible data draws. Thus, the main difficulty in learning $p(\boldsymbol{n})$ is the high-dimensionality of the data space. For example, a 4-second segment of LIGO data sampled at 4096 Hz results in a vector of size 16,384; consequently, the noise distribution $p(\boldsymbol{n})$ is of dimension 16,384. The curse of dimensionality thus hinders the learning process even for flexible models like normalizing flows \citep{2019NFreview, 2019arXiv191202762P}.

An alternative to modeling $p(\boldsymbol{n})$ directly is to learn the \emph{score} of the noise distribution, defined as $\nabla_{\boldsymbol{n}} \log p(\boldsymbol{n})$. This is an easier task, as $\nabla_{\boldsymbol{n}} \log p(\boldsymbol{n})$ is a local quantity independent of the normalization of $p(\boldsymbol{n})$ (which is intractable in high-dimensions), making it possible to learn using any basic function estimator \citep[e.g., simple feedforward neural networks,][]{2019arXiv190705600S}.

As shown in \cite{legin}, we can use the chain rule to relate the score of the noise distribution to the gradient of the log likelihood, which can itself be written as $\nabla_{\boldsymbol{\theta}} \log p(\boldsymbol{d} -\boldsymbol{h}(\boldsymbol{\theta}))$ per Eq.~\eqref{eq:likelihood}. Concretely,
\begin{equation}
  \label{eq:slic}
  \nabla_{\boldsymbol{\theta}} \log p(\boldsymbol{d} - \boldsymbol{h}(\boldsymbol{\theta})) = -\nabla_{\boldsymbol{n}} \log p(\boldsymbol{n})\, \nabla_{\boldsymbol{\theta}} \boldsymbol{h}(\boldsymbol{\theta})\, , 
\end{equation}
where $\nabla_{\boldsymbol{\theta}} \boldsymbol{h}(\boldsymbol{\theta})$ is the Jacobian of the forward model. By combining Eq.~\eqref{eq:slic} with the score of the prior distribution, $\nabla_{\boldsymbol{\theta}} \log p(\boldsymbol{\theta})$, we can obtain the score of the posterior distribution, $\nabla_{\boldsymbol{\theta}} \log p(\boldsymbol{\theta}|\boldsymbol{d})$, via Bayes's theorem.

To sample from the posterior distribution, we utilize gradient-based samplers that leverage the posterior score. In this work, we choose the Metropolis-adjusted Langevin algorithm \citep[MALA,][]{MALA}. As with other Metropolis-Hastings algorithms, MALA involves a proposal step, whereby a new position in the sampling space is proposed, followed by an acceptance/rejection step. For MALA, the proposal only requires the score of the posterior, and not the posterior itself; the rejection step, however, requires knowledge of the posterior probability ratio between the proposed ($\boldsymbol{\theta}_{*}$) and current ($\boldsymbol{\theta}_{n}$) positions.
Following \cite{2023Remy}, the relative probability between two points $\boldsymbol{x}_{*}$ and $\boldsymbol{x}_{n}$ drawn from a distribution $p(\boldsymbol{x})$ can be obtained from the score by solving the line integral
{\small
\begin{equation}
    \label{eq:delta_logp}
    \log \frac{p(\boldsymbol{x}_{*})}{p(\boldsymbol{x}_{n})} = \int_0^1 \nabla \log p(\lambda(\boldsymbol{x}_{*} - \boldsymbol{x}_n) + \boldsymbol{x}_n) \cdot (\boldsymbol{x}_{*} - \boldsymbol{x}_n) \, {\rm d}\lambda\, .
\end{equation}
}We compute the integral in Eq.~\eqref{eq:delta_logp} to approximate the relative posterior probability required for the rejection step (Sec.~\ref{sec:inference}); this can be done to arbitrary precision at the increased computational cost of more score evaluations.

All that remains is to learn the score of the noise distribution, $\nabla_{\boldsymbol{n}} \log p(\boldsymbol{n})$. Next, we detail how we do this using score-generative models, specifically denoising diffusion models, under the SLIC framework \citep{legin,Legin2023icml}.

\subsection{Score Modeling}
\label{sec:score_modeling}

Score-generative models learn the score of a probability distribution, $\nabla_{\boldsymbol{x}} \log p(\boldsymbol{x})$, from a finite set of data draws $\boldsymbol{x}$ \citep{Hyvarinen2005, song2020sde, Ho2020}. In denoising diffusion models \citep{2019arXiv190705600S, song2020sde, Ho2020, 2022arXiv220900796Y}, a model, typically a neural network, is trained to predict the score of a convolved distribution, $\nabla_{\boldsymbol{x}} \log p_{t}(\boldsymbol{x})$, where $t$ is a parameter related to diffusion temperature. For our application, $p_{t}(\boldsymbol{x})$ represents the underlying data distribution $p(\boldsymbol{x})$ convolved with a Gaussian distribution such that the true distribution $p(\boldsymbol{x})$ is recovered as $t \to 0$, i.e., $p_0(\boldsymbol{x}) = p(\boldsymbol{x})$.

Training a neural network to predict $\nabla_{\boldsymbol{x}} \log p_{t}(\boldsymbol{x})$ is accomplished by minimizing the denoising score matching loss \citep{Vincent2011, song2020sde}. The objective is to minimize the difference between the network's output and the score of a perturbation kernel, $\nabla_{\boldsymbol{x}_t} \log p(\boldsymbol{x}_t | \boldsymbol{x}_0)$, with $\boldsymbol{x}_0$ and $\boldsymbol{x}_t$ respectively defined as clean and diffusion-perturbed data. Once trained, we can evaluate the network at $t = 0$, in principle giving us the score of the true data distribution $\nabla_{\boldsymbol{x}} \log p_0(\boldsymbol{x}) = \nabla_{\boldsymbol{x}} \log p(\boldsymbol{x})$.
In practice, learning the true score at $t \to 0$ is challenging and requires substantial training time and careful hyperparameter tuning. Therefore, we approximate the score of the true LIGO noise distribution by setting $t$ to a small but non-zero value (see Section \ref{sec:inference} for details). The following section describes the data used for training the neural network in this paper.

\section{Data}
\label{sec:data}

We train the neural network on 4-second segments of real LIGO noise from both the Livingston and Hanford detectors, obtained from the Gravitational Wave Open Science Center \citep{LIGOScientific:2019lzm,gwosc:bulk}. The training set is composed of data spanning a 4-day window around the first GW detection, GW150914 (GPS time 1126259462.423 s). Additionally, we build a test set using data collected 2 months after the training set. Both data sets are sampled at 4096 Hz. 

Initially, we split the data into 8-second segments, ensuring the segment containing the GW150914 signal is discarded. We then apply a Tukey window with a shape parameter of \(\alpha_T = 0.2\) to each segment to take them to the Fourier domain. The segments are then whitened by subtracting the mean of each Fourier mode and dividing by its standard deviation as estimated over the entire training set. We exclude certain segments contaminated by high signal-to-noise ratio glitches in the computation of the mean and standard deviation; these outlier segments, however, are still included as noise samples in the training set. We finally transform the segments back to the time domain, and crop the boundaries to retain the central 4 seconds of each segment. This conditioning facilitates the training of the network; the details are not essential as long as the same procedure is applied in production.

Our training and testing sets consist of approximately 40,000 and 400 data segments respectively, with half of the segments taken from each of the LIGO detectors. For the purposes of this demonstration, we treat data from individual detectors as independent samples and do not leverage coherence across the network (i.e., we treat all observations as single-detector). In the following section, we detail the neural network's architecture and training procedure.

\section{Neural Network Architecture and Training}
\label{sec:neural_network}

Our neural network follows a U-Net architecture similar to the work of \cite{song2020sde}. Specifically, the network is composed of 8 downsampling/upsampling resolution levels, each with 2 ResNet blocks \citep{Brock2018}. The number of output feature maps from the ResNet blocks for every level is 64, 64, 32, 32, 32, 16, 16, 16. The data is downsampled/upsampled after each level by a factor of  2, 2, 2, 4, 4, 4, 4. During training, we use the Adam optimizer \citep{kingma2014adam} with a learning rate of \(2 \times 10^{-5}\) and a batch size of 32. We apply gradient clipping to restrict the gradients to a maximum global norm of 1. Each ResNet block incorporates Dropout layers \citep{Dropout2014} with a dropout rate set to 10\%. We train the neural network for 500 epochs, equivalent to a training duration of ${\sim}24$ h on a single NVIDIA A100 GPU.

The neural network is trained using the variance-exploding stochastic differential equation (VESDE) model from \cite{song2020sde}. This SDE models a stochastic process where the diffusion noise variance increases as time $t$ progresses from 0 to 1. As such, the $t$ parameter in the convolved score $\nabla_{\boldsymbol{n}} \log p_t(\boldsymbol{n})$ is directly related to the the time evolution of the SDE. In our setup, the distribution $p_{t}(\boldsymbol{n})$ models the LIGO noise, $p(\boldsymbol{n})$, convolved with a Gaussian whose standard deviation $\sigma(t)$ varies with $t$ as
\begin{equation}
    \sigma(t) = \sigma_{\text{min}} \left(\frac{\sigma_{\text{max}}}{\sigma_{\text{min}}}\right)^t,
\end{equation}
where $\sigma_{\text{max}}$ and $\sigma_{\text{min}}$ are hyperparameters that define the maximum and minimum standard deviation. Following the geometric interpretation of \cite{song2020sde}, we set $\sigma_{\text{max}} = 10000$ and $\sigma_{\text{min}} = 0.001$. Our implementation of the neural network and training is publicly available online \citep{scoregen_jax}.

\begin{figure*}[p]
    \centering
    \includegraphics[width=0.95\textwidth]{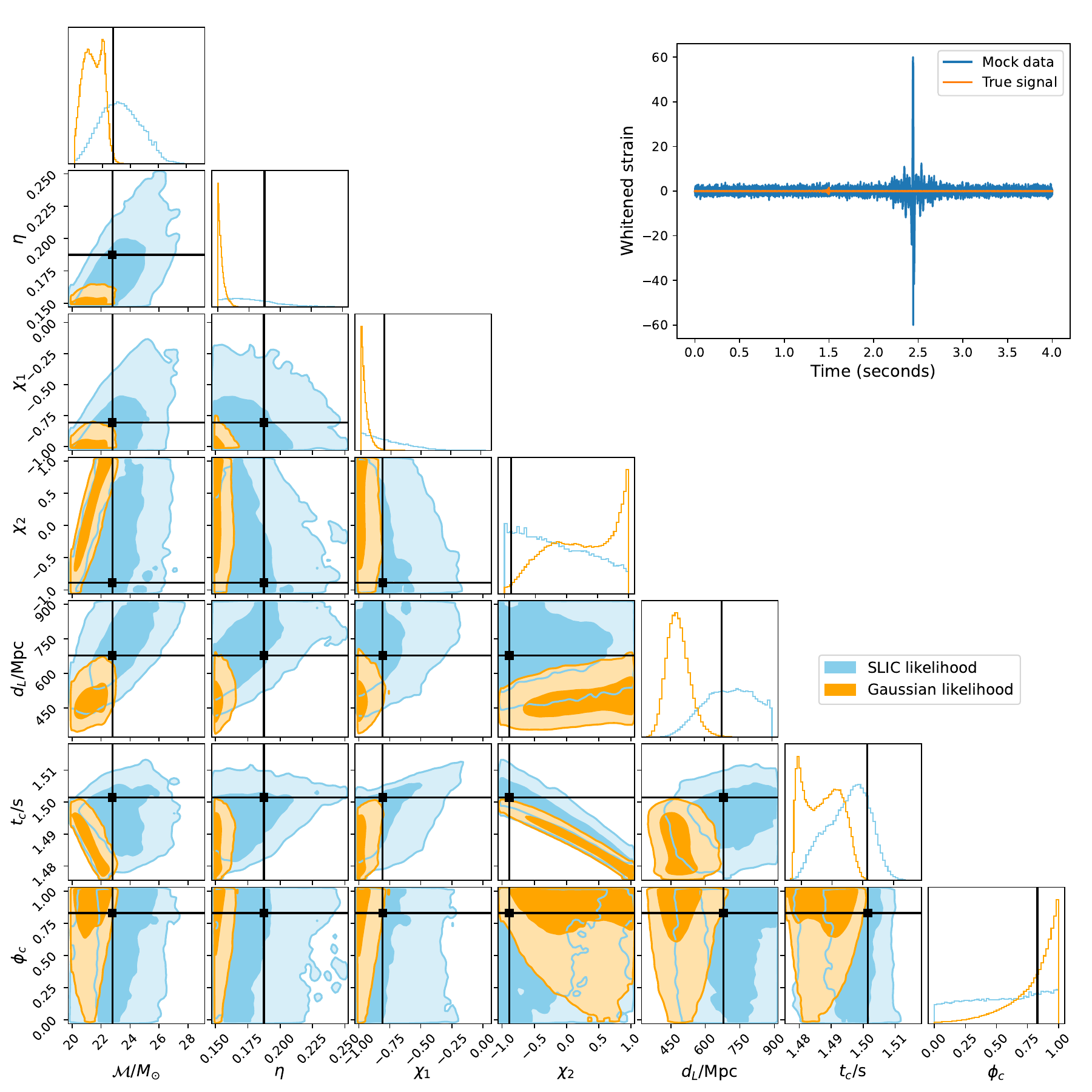}
    \caption{Example of a sampled posterior distribution using SLIC (blue) versus using a Gaussian likelihood (orange). The ground truth model parameter is located at the intersection of the black lines. In the top right corner, the true signal is shown in orange superposed on top of the mock observation in blue. The contours represent the 68\% and 99\% credible intervals. The ground truth parameters of the waveform are $\mathcal{M} = 22.8 M_{\odot}$, $\eta = 0.187$, $\chi_{1} = -0.807$, $\chi_{2} = -0.889$, $d_{L} = 678$ Mpc, and $\phi_{c} = 0.830$. The true trigger time, $t_c$, in relative and absolute GPS time is 1.501 s and 1128689443.501 s, respectively. The LIGO noise segment used in this mock observation starts at an absolute GPS time of 1128689442 s and lasts for 4 seconds. The signal-to-noise ratio of the ground truth waveform is 11.7.}
    \label{fig:corner}
\end{figure*}

\section{Forward Model}
\label{sec:waveform}

For our forward model $\boldsymbol{h}(\boldsymbol{\theta})$, we use the differentiable version of the \textsc{IMRPhenomD} waveform model \citep{Khan:2015jqa} implemented in the \textsc{ripple} package \citep{Edwards:2023sak}.
The ability to automatically differentiate the waveform is crucial to compute the required Jacobian $\nabla_{\boldsymbol{\theta}} \boldsymbol{h}(\boldsymbol{\theta})$ in Eq.~\eqref{eq:slic}. For a binary system with masses $m_{1/2}$ and dimensionless spin magnitudes $\chi_{1/2}$, this waveform model is parameterized in terms of the chirp mass, $\mathcal{M} = (m_1 m_2)^{3/5} / (m_1 + m_2)^{1/5}$, the symmetric mass ratio, $\eta = m_1 m_2 / (m_1 + m_2)$, and the spin magnitudes; the spins are assumed to be colinear with the orbital angular momentum of the binary.

To test SLIC for inference, we sample in $\mathcal{M}$ and $\mathcal{\eta}$, as well as in the source luminosity distance, $d_L$, the coalescence (signal arrival) time, $t_c$, the dimensionless spins, $\chi_{1}$, $\chi_{2}$, and a fiducial coalescence phase, $\phi_c$. To simplify the problem, we assume a sky location given by right ascension $\alpha = 1.95\, \text{rad}$, declination $\delta = -1.27\, \text{rad}$ and polarization angle $\psi = 0.82\, \text{rad}$, consistent with GW150914 \citep{LIGOScientific:2016aoc,LIGOScientific:2016vlm}; also measuring these extrinsic parameters is a straightforward generalization of our setup.

We generate mock observations by simulating GW signals with true parameters randomly sampled from a uniform distribution, which we also adopt as a prior: \(\mathcal{M}/M_\odot \sim \mathcal{U}(20, 40)\), \(\eta \sim \mathcal{U}(0.15, 0.25)\), \(\chi_1, \chi_2 \sim \mathcal{U}(-1.0, 1.0)\), \(d_{L}/\mathrm{Mpc} \sim \mathcal{U}(200, 900)\), \(t_c / \mathrm{s} \sim \mathcal{U}(0.5, 3.5)\), and \(\phi_c \sim \mathcal{U}(0, 1.0)\). The starting frequency for the waveform generator is set to 20 Hz. We add real LIGO Livingston and Hanford detector noise from the test set (see Section \ref{sec:data}) to the simulated signals, generating a total of 200 mock observations with noise from the Livingston detector and 200 from the Hanford detector. Currently, we perform inference separately for each detector; future work will test the SLIC framework using combined data from both detectors, which should improve performance thanks to signal coherence.

\section{Inference}
\label{sec:inference}

To sample from the posterior, we use the network's prediction of the convolved score, $\nabla_{\boldsymbol{n}} \log p_{t}(\boldsymbol{n})$, at $t = 0.3$ (since $t =0$ is numerically unstable; see Sec.~\ref{sec:slic}). Within the settings of our score-generative model, $t = 0.3$ corresponds to the base LIGO noise distribution $p(\boldsymbol{n})$ convolved with a Gaussian with standard deviation of $\sigma(t) = 0.1$, so that the network assumes a baseline noise standard deviation that is ${\sim}0.5\%$ higher than that of the reference PSD. We choose $t = 0.3$ as a safe threshold because our current network becomes evidently unstable for lower $t$ values, yielding spurious score values that result in posteriors significantly more biased than could be accommodated by the associated level of tempering (as apparent from inspection of individual posteriors, or statistical coverage tests). For the rejection step, we approximate Eq.~\eqref{eq:delta_logp} using Simpson's integration rule and discretizing the line integral in two steps. 

As for sampling settings, we initialize 32 walkers in a small region around the ground truth $\boldsymbol{\theta}_{\text{true}}$ and run a warm-up phase of 10,000 steps, during which we adapt the step size and mass matrix for optimal sampling efficiency. After the warm-up phase, we fix the step size and mass matrix and run the chain for an additional 20,000 steps. Samples generated during the warm-up are discarded from the final sampled distribution. 

\begin{figure*}[p]
    \centering
    \includegraphics[width=0.95\textwidth]{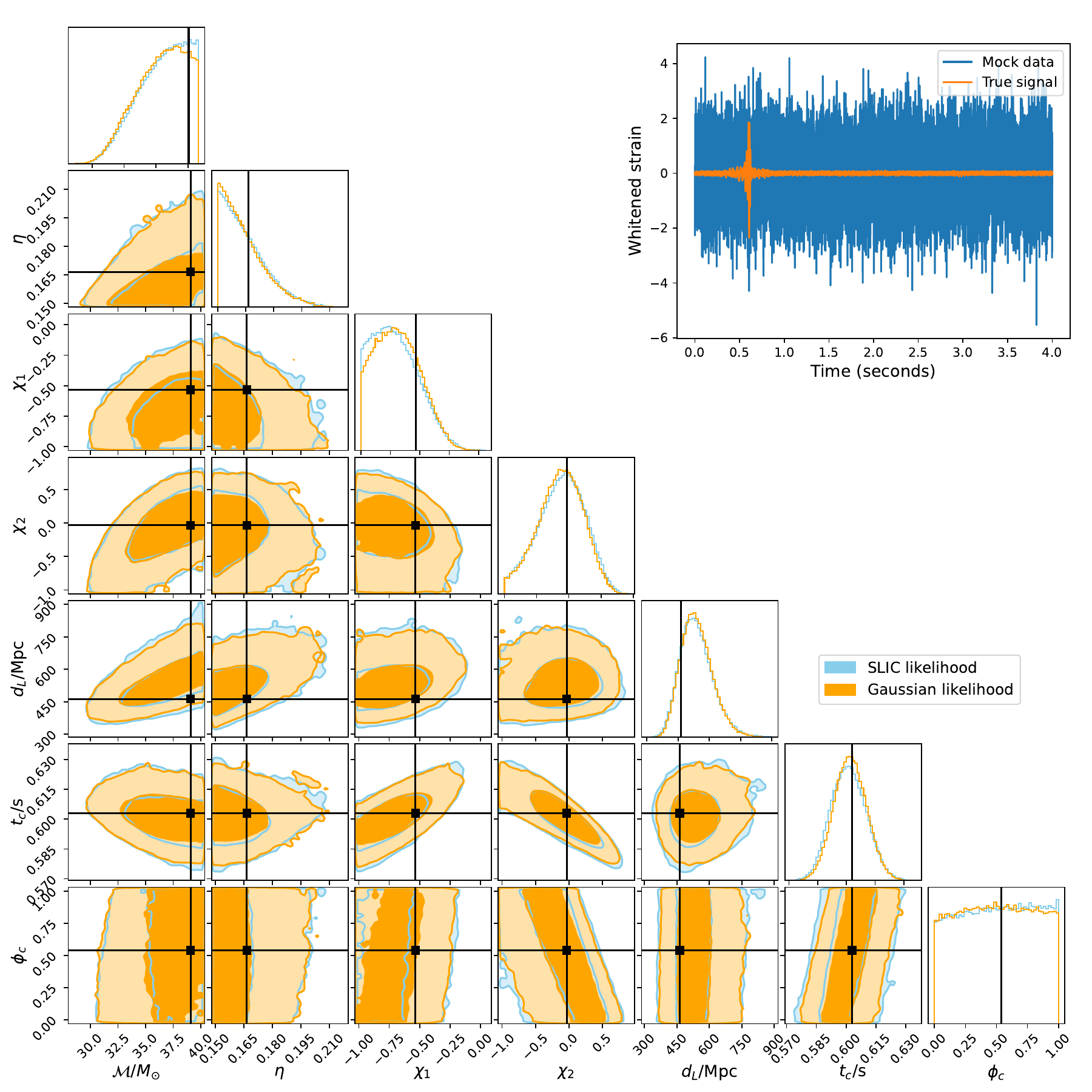}
    \caption{Example similar to Figure \ref{fig:corner} of a posterior distribution sampled using SLIC (blue) compared to a Gaussian likelihood (orange), utilizing real noise from the Livingston detector with no visible signs of non-Gaussianity. The contours indicate the 68\% and 99\% credible intervals. The ground truth parameters for the waveform are: $\mathcal{M} = 39.1 M_{\odot}$, $\eta = 0.167$, $\chi_{1} = -0.534$, $\chi_{2} = -0.033$, $d_{L} = 463$ Mpc, and $\phi_{c} = 0.540$. The true trigger time, $t_c$, is 0.603 s in relative time, corresponding to an absolute GPS time of 1131113090.603 s. The segment of LIGO noise used for this mock observation begins at GPS time 1131113090 s and lasts for 4 seconds. The signal-to-noise ratio of the true waveform is 15.3.}
    \label{fig:corner_gaussian_noise}
\end{figure*}

\section{Results}

\label{sec:results}

In Figure \ref{fig:corner}, we present a mock observation of a simulated signal added to real Hanford data with a loud non-Gaussian glitch. Using the SLIC framework, we obtain a posterior consistent with the ground truth. On the other hand, the standard Gaussian likelihood fails to correctly recover the truth.\footnote{The Gaussian likelihood was computed based on the same whitened residuals used by SLIC.} This illustrates that, by learning the true noise distribution from LIGO detectors, SLIC can achieve accurate inference even when the data are contaminated by strong non-Gaussian features.

In a different setting, Figure \ref{fig:corner_gaussian_noise} compares posterior distributions sampled from a mock observation containing real Livingston noise without visible non-Gaussian features. For this case where the noise more closely follows a Gaussian distribution, the SLIC likelihood returns a result effectively identical to the Gaussian likelihood. Although SLIC is slower than the Gaussian likelihood due to the need for a forward pass through a neural network at each evaluation, it still performs reasonably fast; for instance, running 30,000 MALA steps with SLIC (as for Figs.~\ref{fig:corner} and \ref{fig:corner_gaussian_noise}) takes approximately 1 hour.

\begin{figure}[tb]
    \centering
    \includegraphics[width=\columnwidth]{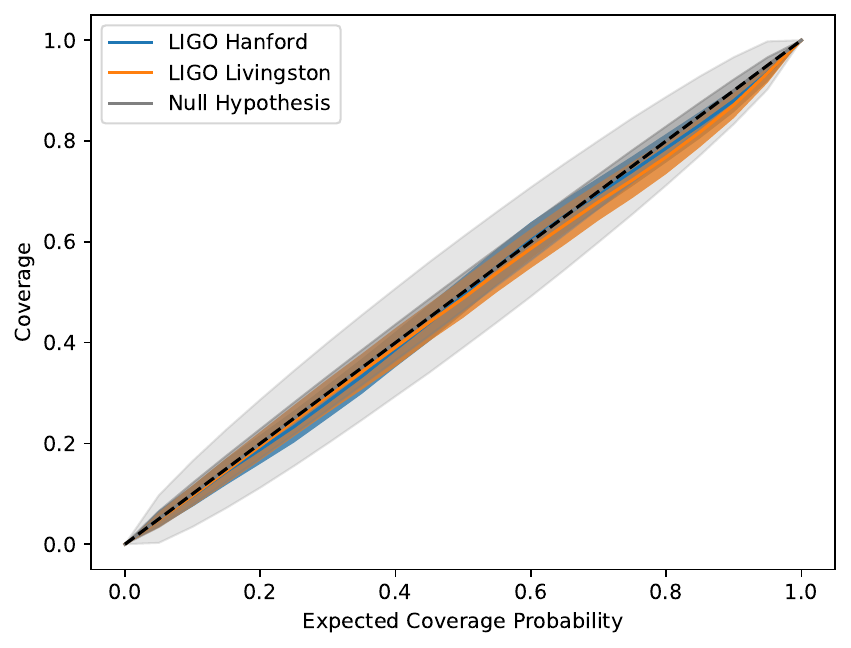}
    \caption{Coverage probability test for our sampled posteriors from 200 mock observations with real noise from the Livingston detector and 200 from the Hanford detector. The orange (Livingston) and blue (Hanford) shaded regions represent the $1 \sigma$ error estimated by bootstrapping over the set of sampled posteriors. The shaded grey regions show the $1 \sigma$ and $3 \sigma$ error of the null hypothesis in which the sampled posteriors are accurate.}
    \label{fig:coverage}
\end{figure}

To more globally assess the accuracy of SLIC, Figure \ref{fig:coverage} shows a probability-probability (PP) coverage plot \citep{cook2006validation,2018arXiv180406788T}, constructed by sampling 400 posterior distributions from our set of 400 mock LIGO observations. We implement this through the test of accuracy with random points (TARP) method, as outlined in \cite{2023Pablo}; TARP verifies whether the ground truth lies within a region that encompasses \(x\%\) of the total posterior probability volume \(x\%\) of the time. As shown in Figure \ref{fig:coverage}, our posterior inference is consistent with the diagonal line, yielding no evidence of bias.  
Code to reproduce our results is openly available online \citep{slicgw_repo}.

\section{Discussion \& Conclusion}
\label{sec:discussion}

The results above demonstrate SLIC's ability to infer the properties of GW sources without bias, even in the presence of strong non-Gaussianities and non-stationarities.
Crucially, it achieves so (1) without requiring any bespoke glitch mitigation treatments and (2) without giving up deterministic waveform models or flexibility in prior choices during sampling.

For simplicity, this demonstration treated all observations as single-detector. This means that the setup here likely underperforms a realistic analysis in which a multi-detector network would help better distinguish coherent signals from incoherent noise.
On the other hand, we have fixed extrinsic source parameters, which would not normally be known \textit{a priori} and would add uncertainty to the measurement; some of this uncertainty would, in turn, be reduced by having multiple detectors.
This does not fundamentally alter our conclusions regarding SLIC's potential.

Our simplified, single-detector setup serves to highlight SLIC's ability to naturally deal with multiple detector configurations, effectively marginalizing over instrumental noise properties.
By pooling data segments from both LIGO Hanford and Livingston in our training set, SLIC simultaneously learned a bimodal noise distribution made up of draws from each detector; as a result, SLIC was able to analyze data from either detector without information about the provenance of any given data segment (i.e., without being told which instrument recorded the data).
Training detector-specific SLIC networks is straightforward and can only improve performance.
SLIC's ability to marginalize over the noise statistics is valuable even in the case of purely Gaussian noise, since the true PSD of the noise is unknown and SLIC effectively marginalizes over it.

Another technical limitation of our current SLIC setup, is the requirement to evaluate the network at $t = 0.3$ rather than $t = 0$ due to numerical stability.
As a consequence, our current implementation of the SLIC likelihood assumes a slightly higher standard deviation in the baseline noise than observed in samples from actual LIGO data (${\sim}0.5\%$ higher than the reference PSD).
In principle, this nonzero annealing of the likelihood results in very slightly overinflated credible regions; however, this is a minuscule effect (as evident from Fig.~\ref{fig:corner_gaussian_noise}) and is not the reason behind the lack of bias when analyzing glitchy data (e.g., the Gaussian likelihood in Fig.~\ref{fig:corner} would still be significantly biased if it was annealed by 0.5\%).
The small reduction in signal-to-noise ratio associated with assuming $t=0.3$ should only become noticeable at the 90\%-credible level for a population analysis of at least ${\sim}10^3$ sources.
In any case, we expect this to not be a fundamental limitation and are currently exploring different network structures and training schemes to be able to lower $t$ as needed.

One of SLIC's benefits compared to other machine learning methods, is its ability to combine the learned noise distribution with any waveform model per Eq.~\eqref{eq:slic}. In contrast, simulation-based inference methods, such as neural posterior estimation (NPE), use neural networks to directly predict the posterior distribution, implicitly encoding the forward model within them \citep{2021PhRvL.127x1103D, 2024PhRvD.109f4056L,Xiong:2024gpx,Raymond:2024xzj}. One significant drawback of these methods is that if the forward model changes, a new neural network must be trained with simulations generated from the new forward model. Additionally, NPEs implicitly learn the prior distribution of model parameters from the training data, making it impossible to update the sampling prior distribution without retraining the neural network. In contrast, SLIC only learns the noise distribution, which is independent of the forward model and, thus, independent of the prior distribution.

In this work, we used MALA as our posterior sampler. When considering higher dimensional generalizations of this problem with nontrivial posterior structure (e.g., to measure the extrinsic parameters), it would likely be advantageous to consider other samplers (such as Hamiltonian Monte Carlo) and better proposal distributions. Integrating the likelihood as learned by our neural network with more powerful sampling algorithms such as \textsc{flowMC} \citep{Wong:2022xvh} promises significant improvements in performance. Such methods, as implemented in \textsc{jim} \citep[e.g.,][]{Wong:2023lgb}, can explore complex, high-dimensional spaces much more efficiently than MALA by optimizing the proposal distribution for sampling efficiency. By combining the SLIC likelihood with an advanced sampler, we could achieve fast, accurate parameter estimation for real gravitational wave events even in the presence of non-Gaussianities and non-stationarity in the data.
In future work, we aim to extend the proposed methodology in this direction.

\section*{Acknowledgements}
We thank Stephen Green for comments on the manuscript, and Colm Talbot for helpful discussions.
This work is partially supported by Schmidt Futures, a philanthropic initiative founded by Eric and Wendy Schmidt as part of the Virtual Institute for Astrophysics (VIA). The Flatiron Institute is supported by the Simons Foundation. The work is in part supported by computational resources provided by Calcul Quebec and the Digital Research Alliance of Canada. Y.H. and L.P. acknowledge support from the Canada Research Chairs Program, the National Sciences and Engineering Council of Canada through grants RGPIN-2020-05073 and 05102, and the Fonds de recherche du Québec through grants 2022-NC-301305 and 300397.
This research has made use of data or software obtained from the Gravitational Wave Open Science Center (gwosc.org), a service of the LIGO Scientific Collaboration, the Virgo Collaboration, and KAGRA. This material is based upon work supported by NSF's LIGO Laboratory which is a major facility fully funded by the National Science Foundation, as well as the Science and Technology Facilities Council (STFC) of the United Kingdom, the Max-Planck-Society (MPS), and the State of Niedersachsen/Germany for support of the construction of Advanced LIGO and construction and operation of the GEO600 detector. Additional support for Advanced LIGO was provided by the Australian Research Council. Virgo is funded, through the European Gravitational Observatory (EGO), by the French Centre National de Recherche Scientifique (CNRS), the Italian Istituto Nazionale di Fisica Nucleare (INFN) and the Dutch Nikhef, with contributions by institutions from Belgium, Germany, Greece, Hungary, Ireland, Japan, Monaco, Poland, Portugal, Spain. KAGRA is supported by Ministry of Education, Culture, Sports, Science and Technology (MEXT), Japan Society for the Promotion of Science (JSPS) in Japan; National Research Foundation (NRF) and Ministry of Science and ICT (MSIT) in Korea; Academia Sinica (AS) and National Science and Technology Council (NSTC) in Taiwan.
This paper carries LIGO document number LIGO-P2400440.

\bibliography{bibliography}{}

\begin{thebibliography}{}
\expandafter\ifx\csname natexlab\endcsname\relax\def\natexlab#1{#1}\fi
\providecommand{\url}[1]{\href{#1}{#1}}
\providecommand{\dodoi}[1]{doi:~\href{http://doi.org/#1}{\nolinkurl{#1}}}
\providecommand{\doeprint}[1]{\href{http://ascl.net/#1}{\nolinkurl{http://ascl.net/#1}}}
\providecommand{\doarXiv}[1]{\href{https://arxiv.org/abs/#1}{\nolinkurl{https://arxiv.org/abs/#1}}}

\bibitem[{Aasi {et~al.}(2015)}]{LIGOScientific:2014pky}
Aasi, J., {et~al.} 2015, Class. Quant. Grav., 32, 074001, \dodoi{10.1088/0264-9381/32/7/074001}

\bibitem[{Abbot {et~al.}(2023)}]{gwosc:bulk}
Abbot, R., {et~al.} 2023, Gravitational Wave Open Science Center (1126216262-1126302662).
\newblock \url{https://gwosc.org/archive/links/O1/H1/1126216262/1126302662/simple/}

\bibitem[{Abbott {et~al.}(2016{\natexlab{a}})}]{LIGOScientific:2016aoc}
Abbott, B.~P., {et~al.} 2016{\natexlab{a}}, Phys. Rev. Lett., 116, 061102, \dodoi{10.1103/PhysRevLett.116.061102}

\bibitem[{Abbott {et~al.}(2016{\natexlab{b}})}]{LIGOScientific:2016gtq}
---. 2016{\natexlab{b}}, Class. Quant. Grav., 33, 134001, \dodoi{10.1088/0264-9381/33/13/134001}

\bibitem[{Abbott {et~al.}(2016{\natexlab{c}})}]{Abbott:2016xvh}
---. 2016{\natexlab{c}}, Phys. Rev. D, 93, 112004, \dodoi{10.1103/PhysRevD.93.112004}

\bibitem[{Abbott {et~al.}(2016{\natexlab{d}})}]{LIGOScientific:2016vlm}
---. 2016{\natexlab{d}}, Phys. Rev. Lett., 116, 241102, \dodoi{10.1103/PhysRevLett.116.241102}

\bibitem[{Abbott {et~al.}(2017)}]{LIGOScientific:2017vwq}
---. 2017, Phys. Rev. Lett., 119, 161101, \dodoi{10.1103/PhysRevLett.119.161101}

\bibitem[{Abbott {et~al.}(2020)}]{LIGOScientific:2019hgc}
---. 2020, Class. Quant. Grav., 37, 055002, \dodoi{10.1088/1361-6382/ab685e}

\bibitem[{Abbott {et~al.}(2021)}]{LIGOScientific:2019lzm}
Abbott, R., {et~al.} 2021, SoftwareX, 13, 100658, \dodoi{10.1016/j.softx.2021.100658}

\bibitem[{Abbott {et~al.}(2023)}]{LIGOScientific:2021djp}
---. 2023, Phys. Rev. X, 13, 041039, \dodoi{10.1103/PhysRevX.13.041039}

\bibitem[{Abbott {et~al.}(2024)}]{LIGOScientific:2021usb}
---. 2024, Phys. Rev. D, 109, 022001, \dodoi{10.1103/PhysRevD.109.022001}

\bibitem[{Acernese {et~al.}(2015)}]{VIRGO:2014yos}
Acernese, F., {et~al.} 2015, Class. Quant. Grav., 32, 024001, \dodoi{10.1088/0264-9381/32/2/024001}

\bibitem[{Ashton(2023)}]{Ashton:2022ztk}
Ashton, G. 2023, Mon. Not. Roy. Astron. Soc., 520, 2983, \dodoi{10.1093/mnras/stad341}

\bibitem[{Ashton {et~al.}(2019)}]{Ashton:2018jfp}
Ashton, G., {et~al.} 2019, Astrophys. J. Suppl., 241, 27, \dodoi{10.3847/1538-4365/ab06fc}

\bibitem[{Aso {et~al.}(2013)Aso, Michimura, Somiya, Ando, Miyakawa, Sekiguchi, Tatsumi, \& Yamamoto}]{Aso:2013eba}
Aso, Y., Michimura, Y., Somiya, K., {et~al.} 2013, Phys. Rev. D, 88, 043007, \dodoi{10.1103/PhysRevD.88.043007}

\bibitem[{Berry {et~al.}(2015)}]{Berry:2014jja}
Berry, C. P.~L., {et~al.} 2015, Astrophys. J., 804, 114, \dodoi{10.1088/0004-637X/804/2/114}

\bibitem[{Biwer {et~al.}(2019)Biwer, Capano, De, Cabero, Brown, Nitz, \& Raymond}]{Biwer:2018osg}
Biwer, C.~M., Capano, C.~D., De, S., {et~al.} 2019, Publ. Astron. Soc. Pac., 131, 024503, \dodoi{10.1088/1538-3873/aaef0b}

\bibitem[{{Brock} {et~al.}(2018){Brock}, {Donahue}, \& {Simonyan}}]{Brock2018}
{Brock}, A., {Donahue}, J., \& {Simonyan}, K. 2018, arXiv e-prints, arXiv:1809.11096, \dodoi{10.48550/arXiv.1809.11096}

\bibitem[{Buikema {et~al.}(2020)}]{aLIGO:2020wna}
Buikema, A., {et~al.} 2020, Phys. Rev. D, 102, 062003, \dodoi{10.1103/PhysRevD.102.062003}

\bibitem[{Chatziioannou {et~al.}(2021)Chatziioannou, Cornish, Wijngaarden, \& Littenberg}]{Chatziioannou:2021ezd}
Chatziioannou, K., Cornish, N., Wijngaarden, M., \& Littenberg, T.~B. 2021, Phys. Rev. D, 103, 044013, \dodoi{10.1103/PhysRevD.103.044013}

\bibitem[{Cook {et~al.}(2006)Cook, Gelman, \& Rubin}]{cook2006validation}
Cook, S.~R., Gelman, A., \& Rubin, D.~B. 2006, Journal of Computational and Graphical Statistics, 15, 675.
\newblock \url{http://www.jstor.org/stable/27594203}

\bibitem[{Cornish \& Littenberg(2015)}]{Cornish:2014kda}
Cornish, N.~J., \& Littenberg, T.~B. 2015, Class. Quant. Grav., 32, 135012, \dodoi{10.1088/0264-9381/32/13/135012}

\bibitem[{Cornish {et~al.}(2021)Cornish, Littenberg, B\'ecsy, Chatziioannou, Clark, Ghonge, \& Millhouse}]{Cornish:2020dwh}
Cornish, N.~J., Littenberg, T.~B., B\'ecsy, B., {et~al.} 2021, Phys. Rev. D, 103, 044006, \dodoi{10.1103/PhysRevD.103.044006}

\bibitem[{Davis {et~al.}(2022)Davis, Littenberg, Romero-Shaw, Millhouse, McIver, Di~Renzo, \& Ashton}]{Davis:2022ird}
Davis, D., Littenberg, T.~B., Romero-Shaw, I.~M., {et~al.} 2022, Class. Quant. Grav., 39, 245013, \dodoi{10.1088/1361-6382/aca238}

\bibitem[{Davis {et~al.}(2019)Davis, Massinger, Lundgren, Driggers, Urban, \& Nuttall}]{Davis:2018yrz}
Davis, D., Massinger, T.~J., Lundgren, A.~P., {et~al.} 2019, Class. Quant. Grav., 36, 055011, \dodoi{10.1088/1361-6382/ab01c5}

\bibitem[{{Dax} {et~al.}(2021){Dax}, {Green}, {Gair}, {Macke}, {Buonanno}, \& {Sch{\"o}lkopf}}]{2021PhRvL.127x1103D}
{Dax}, M., {Green}, S.~R., {Gair}, J., {et~al.} 2021, \prl, 127, 241103, \dodoi{10.1103/PhysRevLett.127.241103}

\bibitem[{Dax {et~al.}(2024)Dax, Green, Gair, Gupte, P\"urrer, Raymond, Wildberger, Macke, Buonanno, \& Sch\"olkopf}]{Dax:2024mcn}
Dax, M., Green, S.~R., Gair, J., {et~al.} 2024.
\newblock \doarXiv{2407.09602}

\bibitem[{Edwards {et~al.}(2024)Edwards, Wong, Lam, Coogan, Foreman-Mackey, Isi, \& Zimmerman}]{Edwards:2023sak}
Edwards, T. D.~P., Wong, K. W.~K., Lam, K. K.~H., {et~al.} 2024, Phys. Rev. D, 110, 064028, \dodoi{10.1103/PhysRevD.110.064028}

\bibitem[{Hannam {et~al.}(2022)}]{Hannam:2021pit}
Hannam, M., {et~al.} 2022, Nature, 610, 652, \dodoi{10.1038/s41586-022-05212-z}

\bibitem[{Heinzel {et~al.}(2023)Heinzel, Talbot, Ashton, \& Vitale}]{Heinzel:2023vkq}
Heinzel, J., Talbot, C., Ashton, G., \& Vitale, S. 2023, Mon. Not. Roy. Astron. Soc., 523, 5972, \dodoi{10.1093/mnras/stad1823}

\bibitem[{{Ho} {et~al.}(2020){Ho}, {Jain}, \& {Abbeel}}]{Ho2020}
{Ho}, J., {Jain}, A., \& {Abbeel}, P. 2020, arXiv e-prints, arXiv:2006.11239, \dodoi{10.48550/arXiv.2006.11239}

\bibitem[{Hourihane {et~al.}(2022)Hourihane, Chatziioannou, Wijngaarden, Davis, Littenberg, \& Cornish}]{Hourihane:2022doe}
Hourihane, S., Chatziioannou, K., Wijngaarden, M., {et~al.} 2022, Phys. Rev. D, 106, 042006, \dodoi{10.1103/PhysRevD.106.042006}

\bibitem[{Hyv\"{a}rinen(2005)}]{Hyvarinen2005}
Hyv\"{a}rinen, A. 2005, J. Mach. Learn. Res., 6, 695–709

\bibitem[{Khan {et~al.}(2016)Khan, Husa, Hannam, Ohme, P\"urrer, Jim\'enez~Forteza, \& Boh\'e}]{Khan:2015jqa}
Khan, S., Husa, S., Hannam, M., {et~al.} 2016, Phys. Rev. D, 93, 044007, \dodoi{10.1103/PhysRevD.93.044007}

\bibitem[{Kingma \& Ba(2014)}]{kingma2014adam}
Kingma, D.~P., \& Ba, J. 2014, arXiv preprint arXiv:1412.6980

\bibitem[{{Kobyzev} {et~al.}(2019){Kobyzev}, {Prince}, \& {Brubaker}}]{2019NFreview}
{Kobyzev}, I., {Prince}, S. J.~D., \& {Brubaker}, M.~A. 2019, arXiv e-prints, arXiv:1908.09257, \dodoi{10.48550/arXiv.1908.09257}

\bibitem[{Legin(2024)}]{scoregen_jax}
Legin, R. 2024, scoregen\_jax.
\newblock \url{https://github.com/RonanLegin/scoregen_jax}

\bibitem[{{Legin} {et~al.}(2023){Legin}, {Adam}, {Hezaveh}, \& {Perreault Levasseur}}]{legin}
{Legin}, R., {Adam}, A., {Hezaveh}, Y., \& {Perreault Levasseur}, L. 2023, arXiv e-prints, arXiv:2302.03046, \dodoi{10.48550/arXiv.2302.03046}

\bibitem[{Legin {et~al.}(2023)Legin, Isi, Wong, Adam, Perreault-Levasseur, \& Hezaveh}]{Legin2023icml}
Legin, R., Isi, M., Wong, K., {et~al.} 2023, in Machine Learning for Astrophysics. Workshop at the Fortieth International Conference on Machine Learning (ICML 2023), 17.
\newblock \url{https://ml4astro.github.io/icml2023/assets/70.pdf}

\bibitem[{Legin {et~al.}(2024)Legin, Wong, \& Isi}]{slicgw_repo}
Legin, R., Wong, K., \& Isi, M. 2024, slicgw.
\newblock \url{https://github.com/RonanLegin/slicgw}

\bibitem[{{Lemos} {et~al.}(2023){Lemos}, {Coogan}, {Hezaveh}, \& {Perreault-Levasseur}}]{2023Pablo}
{Lemos}, P., {Coogan}, A., {Hezaveh}, Y., \& {Perreault-Levasseur}, L. 2023, 40th International Conference on Machine Learning, 202, 19256, \dodoi{10.48550/arXiv.2302.03026}

\bibitem[{{Leyde} {et~al.}(2024){Leyde}, {Green}, {Toubiana}, \& {Gair}}]{2024PhRvD.109f4056L}
{Leyde}, K., {Green}, S.~R., {Toubiana}, A., \& {Gair}, J. 2024, \prd, 109, 064056, \dodoi{10.1103/PhysRevD.109.064056}

\bibitem[{Macas {et~al.}(2024)Macas, Lundgren, \& Ashton}]{Macas:2023wiw}
Macas, R., Lundgren, A., \& Ashton, G. 2024, Phys. Rev. D, 109, 062006, \dodoi{10.1103/PhysRevD.109.062006}

\bibitem[{Pankow {et~al.}(2018)}]{Pankow:2018qpo}
Pankow, C., {et~al.} 2018, Phys. Rev. D, 98, 084016, \dodoi{10.1103/PhysRevD.98.084016}

\bibitem[{{Papamakarios} {et~al.}(2019){Papamakarios}, {Nalisnick}, {Jimenez Rezende}, {Mohamed}, \& {Lakshminarayanan}}]{2019arXiv191202762P}
{Papamakarios}, G., {Nalisnick}, E., {Jimenez Rezende}, D., {Mohamed}, S., \& {Lakshminarayanan}, B. 2019, arXiv e-prints, arXiv:1912.02762, \dodoi{10.48550/arXiv.1912.02762}

\bibitem[{Payne {et~al.}(2022)Payne, Hourihane, Golomb, Udall, Udall, Davis, \& Chatziioannou}]{Payne:2022spz}
Payne, E., Hourihane, S., Golomb, J., {et~al.} 2022, Phys. Rev. D, 106, 104017, \dodoi{10.1103/PhysRevD.106.104017}

\bibitem[{Raymond {et~al.}(2024)Raymond, Al-Shammari, \& G\"ottel}]{Raymond:2024xzj}
Raymond, V., Al-Shammari, S., \& G\"ottel, A. 2024.
\newblock \doarXiv{2406.03935}

\bibitem[{{Remy} {et~al.}(2023){Remy}, {Lanusse}, {Jeffrey}, {Liu}, {Starck}, {Osato}, \& {Schrabback}}]{2023Remy}
{Remy}, B., {Lanusse}, F., {Jeffrey}, N., {et~al.} 2023, \aap, 672, A51, \dodoi{10.1051/0004-6361/202243054}

\bibitem[{Roberts \& Tweedie(1996)}]{MALA}
Roberts, G.~O., \& Tweedie, R.~L. 1996, Bernoulli, 2, 341.
\newblock \url{http://www.jstor.org/stable/3318418}

\bibitem[{Romero-Shaw {et~al.}(2020)}]{Romero-Shaw:2020owr}
Romero-Shaw, I.~M., {et~al.} 2020, Mon. Not. Roy. Astron. Soc., 499, 3295, \dodoi{10.1093/mnras/staa2850}

\bibitem[{{Song} \& {Ermon}(2019)}]{2019arXiv190705600S}
{Song}, Y., \& {Ermon}, S. 2019, arXiv e-prints, arXiv:1907.05600, \dodoi{10.48550/arXiv.1907.05600}

\bibitem[{{Song} {et~al.}(2020){Song}, {Sohl-Dickstein}, {Kingma}, {Kumar}, {Ermon}, \& {Poole}}]{song2020sde}
{Song}, Y., {Sohl-Dickstein}, J., {Kingma}, D.~P., {et~al.} 2020, arXiv e-prints, arXiv:2011.13456, \dodoi{10.48550/arXiv.2011.13456}

\bibitem[{Srivastava {et~al.}(2014)Srivastava, Hinton, Krizhevsky, Sutskever, \& Salakhutdinov}]{Dropout2014}
Srivastava, N., Hinton, G., Krizhevsky, A., Sutskever, I., \& Salakhutdinov, R. 2014, Journal of Machine Learning Research, 15, 1929.
\newblock \url{http://jmlr.org/papers/v15/srivastava14a.html}

\bibitem[{{Talts} {et~al.}(2018){Talts}, {Betancourt}, {Simpson}, {Vehtari}, \& {Gelman}}]{2018arXiv180406788T}
{Talts}, S., {Betancourt}, M., {Simpson}, D., {Vehtari}, A., \& {Gelman}, A. 2018, arXiv e-prints, arXiv:1804.06788, \dodoi{10.48550/arXiv.1804.06788}

\bibitem[{Udall \& Davis(2023)}]{Udall:2022vkv}
Udall, R., \& Davis, D. 2023, Appl. Phys. Lett., 122, 094103, \dodoi{10.1063/5.0136896}

\bibitem[{Udall {et~al.}(2024)Udall, Hourihane, Miller, Davis, Chatziioannou, Isi, \& Deshong}]{Udall:2024ovp}
Udall, R., Hourihane, S., Miller, S., {et~al.} 2024.
\newblock \doarXiv{2409.03912}

\bibitem[{Veitch {et~al.}(2015)}]{Veitch:2014wba}
Veitch, J., {et~al.} 2015, Phys. Rev. D, 91, 042003, \dodoi{10.1103/PhysRevD.91.042003}

\bibitem[{Vincent(2011)}]{Vincent2011}
Vincent, P. 2011, Neural Comput., 23, 1661, \dodoi{10.1162/NECO\_a\_00142}

\bibitem[{Wong {et~al.}(2023{\natexlab{a}})Wong, Gabri\'e, \& Foreman-Mackey}]{Wong:2022xvh}
Wong, K. W.~k., Gabri\'e, M., \& Foreman-Mackey, D. 2023{\natexlab{a}}, J. Open Source Softw., 8, 5021, \dodoi{10.21105/joss.05021}

\bibitem[{Wong {et~al.}(2023{\natexlab{b}})Wong, Isi, \& Edwards}]{Wong:2023lgb}
Wong, K. W.~K., Isi, M., \& Edwards, T. D.~P. 2023{\natexlab{b}}, Astrophys. J., 958, 129, \dodoi{10.3847/1538-4357/acf5cd}

\bibitem[{Xiong {et~al.}(2024)Xiong, Sun, Zhang, \& Zhang}]{Xiong:2024gpx}
Xiong, C.-Y., Sun, T.-Y., Zhang, J.-F., \& Zhang, X. 2024.
\newblock \doarXiv{2405.09475}

\bibitem[{{Yang} {et~al.}(2022){Yang}, {Zhang}, {Song}, {Hong}, {Xu}, {Zhao}, {Zhang}, {Cui}, \& {Yang}}]{2022arXiv220900796Y}
{Yang}, L., {Zhang}, Z., {Song}, Y., {et~al.} 2022, arXiv e-prints, arXiv:2209.00796, \dodoi{10.48550/arXiv.2209.00796}

\end{thebibliography}
\bibliographystyle{aasjournal}



\end{document}